\documentclass[twocolumn,aps,prl]{revtex4-1}
\newlength\mylen
\usepackage{graphicx}
\usepackage{amsmath}
\usepackage{url}
\usepackage{float}
\usepackage[english]{babel}
\usepackage{chngcntr}
\usepackage{natbib}
\usepackage{textcase}
\usepackage{zref-xr}
\usepackage{hyperref}
\usepackage{tikz}
\usepackage{xcolor}
\usepackage{epstopdf}
\usepackage[percent]{overpic}
\usepackage{chngcntr}
\usepackage{mathtools}
\usepackage{rotating}
\usepackage{placeins}
\usepackage[absolute,overlay]{textpos}
\usepackage{array}
\newcolumntype{C}{>{\hfil$}p{\mylen}<{$\hfil}}

\usepackage[absolute,overlay]{textpos}
\zexternaldocument[B-]{Supplementary}
\hypersetup{
    colorlinks,
    citecolor=blue,
    filecolor=blue,
    linkcolor=blue,
    urlcolor=blue
}

\begin{document}

\title{Low-Error Operation of Spin Qubits with Superexchange Coupling}
\author{Marko J. Ran\v{c}i\'{c}}
\thanks{Current address: Department of Physics, University of Basel, Klingelbergstrasse 82, CH-4056 Basel, Switzerland} 
\email[e-mail: ]{marko.rancic@uni-konstanz.de}
\author{Guido Burkard}
\email[e-mail: ]{guido.burkard@uni-konstanz.de}
\affiliation{Department of Physics, University of Konstanz, D-78457 Konstanz, Germany}

\date{\today}

\begin{abstract}
In this theoretical work we investigate superexchange, as a means of indirect exchange interaction between two single electron spin qubits, each embedded in a single semiconductor quantum dot (QD). The exchange interaction is mediated by an intermediate, empty QD. 
Our findings suggest the existence of first order ``super sweet spots", in which the qubit operations implemented by superexchange interaction are simultaneously insensitive to charge noise and errors due to spin-orbit interaction.
We also find that the sign of the superexchange can be changed by varying the energy detunings between the QDs.

\end{abstract}

\maketitle

\textit{Introduction.--}Noise-insensitive control of qubits is an important task in quantum information science \cite{4Loss1,4Burkard2,4Hu2,4Hanson1}.
 In addition to its use for two-qubit operations of single electron spin qubits \cite{4Petta1}, the exchange interaction has been utilized to control double \cite{4Levy1,4Reilly1,4Wu1,4Petta1} and triple electron spin qubits \cite{4Divincenzo1,4Medford1,4Medford2,4Medford3} in semiconductor quantum dots (QDs).
However, overcoming the sensitivity of exchange interaction to charge noise \cite{4Burkard2,4Hu2} and errors originating from spin-orbit interaction \cite{4Bonesteel1,4Burkard1} has proved to be a challenging task.  

Three electron spin qubits can be operated close to a ``sweet spot", where the sensitivity of exchange interaction to charge noise vanishes in first order \cite{4Medford1,4Medford2,4Medford3,4Russ1}.
On the other hand, two-electron $S-T_0$ spin qubits embedded in double QDs, only have a trivial first order ``sweet spot", where the exchange interaction is smallest ($\sim t^2/U)$. A possibility to reduce the sensitivity of the $S-T_0$ qubit to electric noise is to control the magnitude of the exchange interaction by controlling the tunnel coupling instead 
of controlling the detuning between the two dots (symmetric operation) \cite{4Reed1,4Martins1}.

The spin-orbit interaction represents a powerful resource to control spin qubits \cite{4Nadj1,4Nadj2}. On the other hand, it can also reduce the coherence time of the electron spin qubit, hamper efforts to prolong the coherence time of the electron spin qubit \cite{4Rancic1,4Nichol1}, and lead to errors in two-qubit operations \cite{4Bonesteel1,4Burkard1}.

Superexchange is the underlying mechanism responsible for the creation of antiferomagnetic order in CuO and MnO \cite{kramers1934interaction, PhysRev.79.350}, is a possible mechanism for $d$-wave high $T_c$ superconductivity \cite{PhysRevB.38.5142}, and allows for switching between ferromagnetic and anti-ferromagnetic order in cold atomic gases \cite{trotzky2008time}.
Although the possibility to use mediated exchange (superexchange) was mentioned in the original Loss-DiVincenzo proposal \cite{4Loss1}, superexchange has not received significant attention from the spin qubit community (see, however, refs. \cite{4Trif1, PhysRevB.89.161402,4Baart1}). One of the reasons for this lies in the fact that compared to the direct exchange superexchange requires an additional quantum dot.

\begin{figure}[t!]
	\centering
	\includegraphics[width=0.48\textwidth]{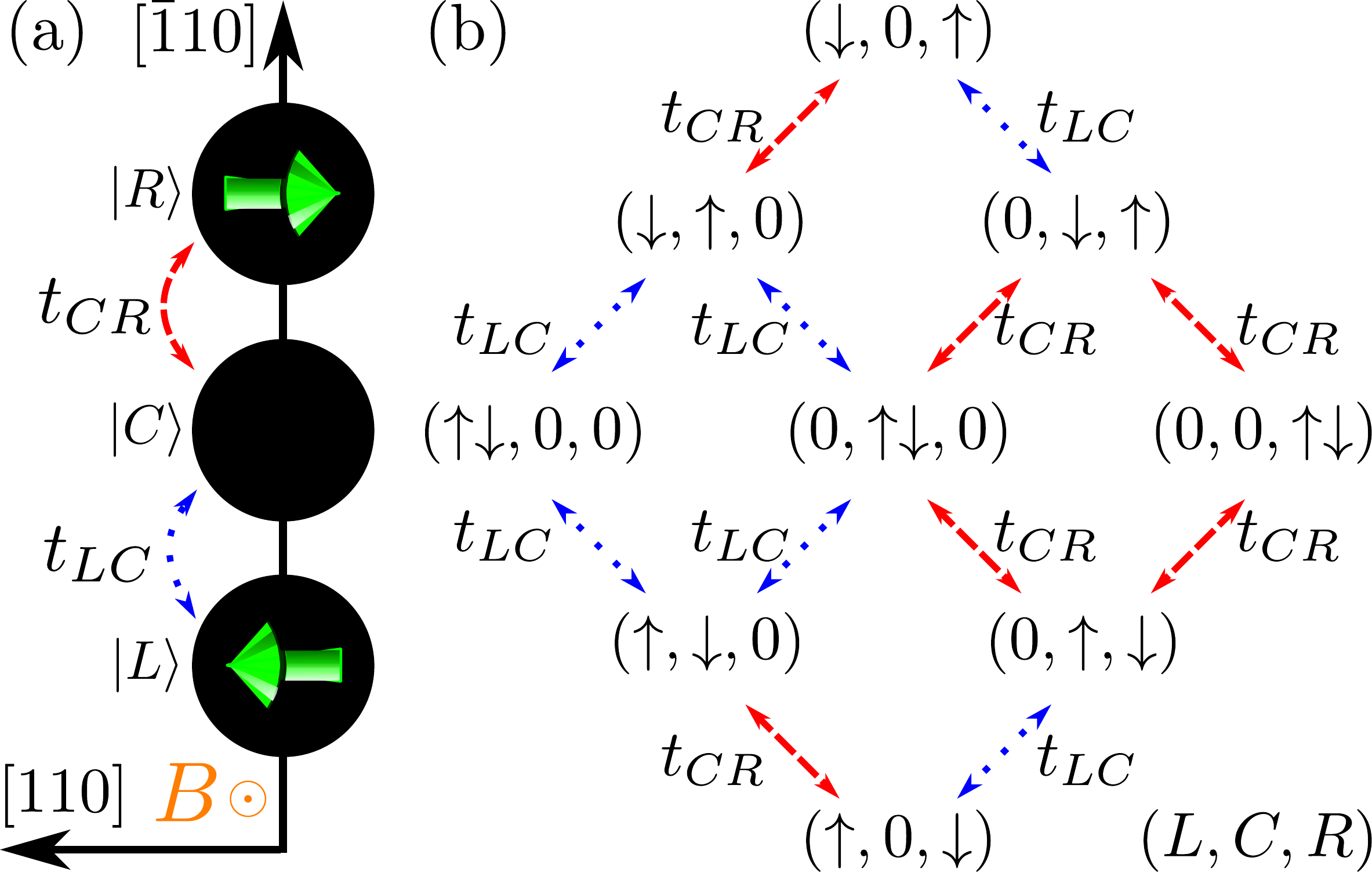}
	\caption{(Color online) (a) The geometry of the system, where $B$ denotes the direction of the external magnetic field, $t_{ LC}$ are spin-conserving hoppings between the left ${(L)}$ and center ${(C)}$ dot (marked with dotted blue lines), $t_{CR}$ are spin-conserving hoppings between the ${C}$ and the right ${ (R)}$ dot (marked with dashed red lines) and $[110]$ and $[\bar{1}10]$ are the crystallographic axes. (b) The scheme of all possible superexchange paths in absence of spin-orbit interactions. 
	All superexchange paths involve four tunneling events, two between the ${ L}$ and the ${C}$ QDs $t_{LC}$ and two between the ${C}$ and the ${R}$ QDs  $t_{CR}$. $\uparrow$ stands for a spin up state, $\downarrow$ for a spin down state and fields in the parentheses denote charge occupancies of the ${(L,C,R)}$ QDs.}
	\label{SupExcSch}
\end{figure}

In this theoretical paper, we investigate superexchange, the exchange interaction between two single electron spin qubits, each embedded in a semiconductor QD on the left $(L)$ and right $(R)$, mediated by an empty quantum dot in the center $(C)$ (see Fig. \ref{SupExcSch} (a)). We have discovered 
a parameter regime in which the superexchange is non-zero and is simultaneously insensitive to both charge noise and errors due to spin-orbit interaction in first order (a non-trivial first order ``super sweet spot"). Our further findings suggest that the sign and the magnitude of superexchange can be controlled by varying the detunings between the QDs.

\textit{Model.--}The superexchange is a fourth-order tunneling process, in which the $(1,0,1)$ charge state with antiparallel spins, virtually tunnels via the 
$(1,1,0)$ or $(0,1,1)$ state to the $(2,0,0)$, $(0,2,0)$ or $(0,0,2)$ charge state, followed by a tunneling back to the $(1,1,0)$ or $(0,1,1)$ state and finally again to the $(1,0,1)$ charge state,
but with the spin state of the $L$ and $R$ QD exchanged, as shown in Fig. \ref{SupExcSch}.

We describe the system with a generalized Hubbard Hamiltonian $H=H_0+H'$ for two electrons in a triple quantum dot,
\begin{align}
 H_0=&\sum\limits_{i\sigma}(\varepsilon_i+E_{\rm z}^i{\bf \sigma})n_{i\sigma}+U\sum\limits_i n_{i\uparrow} n_{i\downarrow}+\sum\limits_{\langle ij \rangle}V n_in_j, \label{eq:Ham0}\\
  H'=&\sum\limits_{\langle ij \rangle}\Big[\sum \limits_{\sigma} t_{ij} c_{i\sigma}^{\dagger}c_{j\sigma}+ \sum \limits_{\sigma\neq\bar{\sigma}} t_{ij}^{\rm so} c_{i\sigma}^{\dagger}c_{j\bar{\sigma}}\Big]. \label{eq:HamPrime}
\end{align}
Here, $E_{\rm z}$ is the Zeeman energy due to an external magnetic field, $t_{ij}$ and $t_{ij}^{\rm so}$ are the magnitudes of spin-conserving and
spin-orbit-induced spin-non-conserving tunnel hoppings respectively, between dots $i$ and $j$. Furthermore, $\varepsilon_i$ denotes the energy bias of the $i$-th dot, $U$ is the Coulomb penalization of the doubly occupied quantum dot, $V$ is the Coulomb energy of two neighboring dots occupied with single electrons and $n_i=n_{i\uparrow}+n_{i\downarrow}=c_{i\uparrow}^{\dagger}c_{i\uparrow}+c_{i\downarrow}^{\dagger}c_{i\downarrow}$ the number operator, with $c_{i\sigma}(c_{i\sigma}^{\dagger})$ being the spin creation (annihilation) operator of the $i$ charge state with spin $\sigma=\downarrow$, $\uparrow$.
The $\langle ij\rangle$ in the index of the sum denotes that the sum runs over nearest neighbor QDs $i$ and $j$, 
and the index $\sigma\neq\bar{\sigma}$ denotes a restricted double sum which runs over all possible states of different spin.

\begin{figure}[t!]
	\centering
	\includegraphics[width=0.35\textwidth]{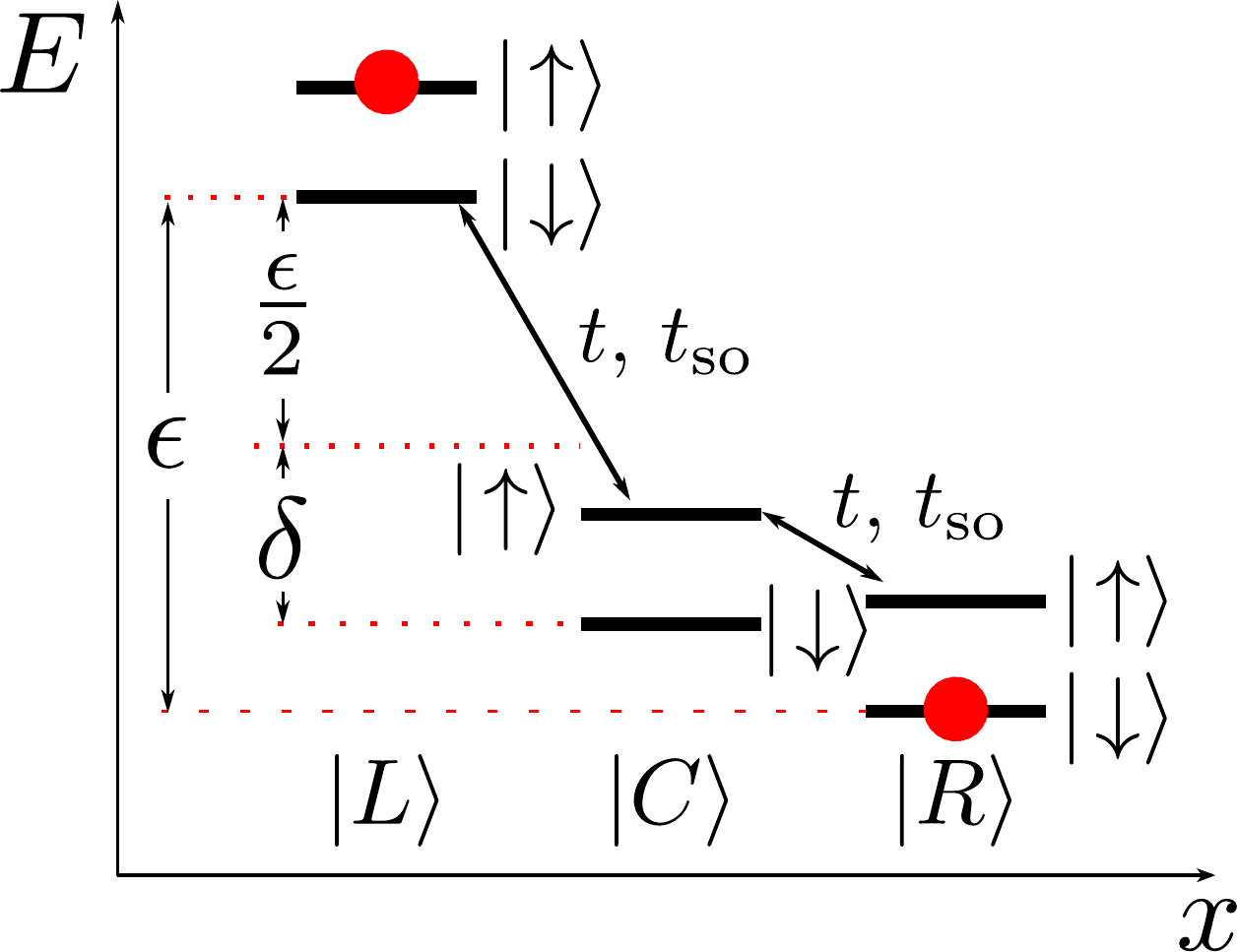}
	\caption{Level diagram, where $E$ denotes the energy, $x$ the position, $\epsilon$ is the energy difference between the outer dots ($L$ and $R$), and $\delta$ the energy between the average energy of the outer dots ($L$ and $R$) and the center $(C)$ QD.}
	\label{EnDig}
\end{figure}

The Coulomb repulsion of doubly occupied quantum dots is characterized by an energy of $U\sim 1\text{ meV}$, and the Coulomb repulsion of neighboring dots being occupied $V\sim0.1U-0.01U$.
Therefore, we neglect the Coulomb repulsion of neighboring dots for simplicity. We also assume a linear triple QD arrangement, allowing us to neglect direct hopping between the R and the L dot, $t_{LR}=t_{LR}^{\rm so}=0$.
Furthermore, from now on we will assume that $t_{LC}=t_{CR}=t$, $t_{LC}^{\rm so}=t_{CR}^{\rm so}=t_{\rm so}$, a 2DEG in the $(001)$ plane of a zincblende semiconductor and Rashba $\alpha$ and Dresselhaus $\beta$ spin-orbit constants of same signs \cite{4Giglberger1}. This means that the magnitude of the spin-orbit hopping $t_{\rm so}$ is maximal when the linear triple quantum dot is structured along the $[\bar{1}10]$ crystallographic axis and minimal when the triple quantum dot is structured along $[110]$ (see Fig. \ref{SupExcSch} (a)). The relation between the spin-conserving $t$ and spin-non-conserving $t_{\rm so}$ hopping is given by $t_{\rm so}=4 t l/3\Lambda_{\rm so},$
where, $l$ is the interdot separation, and $\Lambda_{\rm so}={\hbar/m^*\sqrt{(\alpha+\beta)^2\sin^2{\phi}+(\alpha-\beta)^2\cos^2{\phi}}}$ is the spin-orbit length, where $\phi$ is the angle between the $[110]$ crystallographic axis and the interdot connection axis. Detunings in the Hamiltonian Eq. (\ref{eq:Ham0}) can be expressed in terms of two parameters, the detuning between the outer dots $\epsilon$ and the detuning between the center dot and average detuning of the outer dots $\delta$ Fig. \ref{EnDig}.

\begin{center}
\begin{table*}[t]
\begin{tabular}{|c|c|c|c|}
\hline
 i & Superexchange path & Superexchange expression & Sign of $J_{\rm SE}^i$\\
 \hline

1 & ${(\uparrow,0,\downarrow)\xleftrightarrow[]{t_R}(\uparrow,\downarrow,0)\xleftrightarrow[]{t_L}(0,\uparrow\downarrow,0)\xleftrightarrow[]{t_L}(\downarrow,\uparrow,0)\xleftrightarrow[]{t_R}(\downarrow,0,\uparrow)}$ & $ t^4/\left[(U-2\delta)(\epsilon/2+\delta)^2\right]$  & $J^1_{\rm SE}>0$\\ 
2 & ${(\uparrow,0,\downarrow)\xleftrightarrow[]{t_L}(0,\uparrow,\downarrow)\xleftrightarrow[]{t_R}(0,\uparrow\downarrow,0)\xleftrightarrow[]{t_R}(0,\downarrow,\uparrow)\xleftrightarrow[]{t_L}(\downarrow,0,\uparrow)}$ & $ t^4/\left[(U-2\delta)(\epsilon/2-\delta)^2\right]$ & $J^2_{\rm SE}>0$\\ 
3 & ${(\uparrow,0,\downarrow)\xleftrightarrow[]{t_R}(\uparrow,\downarrow,0)\xleftrightarrow[]{t_L}(0,\uparrow\downarrow,0)\xleftrightarrow[]{t_L}(0,\uparrow,\downarrow)\xleftrightarrow[]{t_L}(\downarrow,0,\uparrow)}$ & $ -t^4/\left[(U-2\delta)(\epsilon/2-\delta)(\epsilon/2+\delta)\right]$ & $J^3_{\rm SE}<0$\\
4 & ${(\uparrow,0,\downarrow)\xleftrightarrow[]{t_L}(0,\uparrow,\downarrow)\xleftrightarrow[]{t_R}(0,\uparrow\downarrow,0)\xleftrightarrow[]{t_L}(\downarrow,\uparrow,0)\xleftrightarrow[]{t_R}(\downarrow,0,\uparrow)}$ & $ -t^4/\left[(U-2\delta)(\epsilon/2-\delta)(\epsilon/2+\delta)\right]$ & $J^4_{\rm SE}<0$\\ 
5 & ${(\uparrow,0,\downarrow)\xleftrightarrow[]{t_R}(\uparrow,\downarrow,0)\xleftrightarrow[]{t_L}(\uparrow\downarrow,0,0)\xleftrightarrow[]{t_L}(\downarrow,\uparrow,0)\xleftrightarrow[]{t_R}(\downarrow,0,\uparrow)}$ & $ t^4/\left[(U-\epsilon)(\epsilon/2+\delta)^2\right]$ & $J^5_{\rm SE}>0$\\ 
6 & ${(\uparrow,0,\downarrow)\xleftrightarrow[]{t_L}(0,\uparrow,\downarrow)\xleftrightarrow[]{t_R}(0,0,\uparrow\downarrow)\xleftrightarrow[]{t_R}(0,\downarrow,\uparrow)\xleftrightarrow[]{t_L}(\downarrow,0,\uparrow)}$ & $ t^4/\left[(U+\epsilon)(\epsilon/2-\delta)^2\right]$ & $J^6_{\rm SE}<0$ \\
\hline
\end{tabular}
\caption{Six possible superexchange paths involving spin-conserving tunneling with corresponding expressions $J_{\rm SE}=\sum_i J_{\rm SE}^i$. The parameters for which the sign of $J_{\rm SE}$ is valid are the Coulomb repulsion $U=1\text{ meV}$, the detuning between the outer dots $\epsilon=-1.34U$, the detuning between the middle dot and the average of the outer dots $-0.2U<\delta<0.3U$.}
\label{tab1}
\end{table*}
\end{center}

\textit{Results.}--We transform the initial generalized Hubbard Hamiltonian $H=H_0+H'$ (see Eq. (\ref{eq:Ham0}) and Eq. (\ref{eq:HamPrime}))
by means of a fourth order Schrieffer-Wolff (SW) transformation, yielding an effective Hamiltonian in which the superexchange subspace $s=\{{(\uparrow,0,\downarrow)},\,{(\downarrow,0,\uparrow),}\,{(\uparrow,0,\uparrow),}\,{(\downarrow,0,\downarrow)}\}$ is decoupled
from the $11$ dimensional subspace of high energy states $h=
\{{(\uparrow,\downarrow,0),\,\,\,(\downarrow,\uparrow,0),\,\,\,(\uparrow,\uparrow,0),\,\,\,(\downarrow,\downarrow,0),\,\,\,(0,\uparrow,\downarrow),\,\,\,(0,\downarrow,\uparrow),}\\
{(0,\uparrow,\uparrow),}\, {(0,\downarrow,\downarrow),}\,{(\uparrow\downarrow,0,0),}\,{(0,\uparrow\downarrow,0),}\,{(0,0,\uparrow\downarrow)}\}$ (for more details about the SW transformation see the Supplementary material). For a linear quantum dot structured along $[\bar{1}10]$ and an external magnetic field parallel to the $(001)$ direction, the effective Hamiltonian up to forth order in perturbation theory in $t_{ij}$ and $t_{ij}^{\rm so}$ within the superexchange subspace $s$ is
\begin{equation}\label{eq:H2x2so}
\tilde{H}=J_{\rm SE}{\bf S}_{ L}\cdot {\bf S}_{R}+D(S_{L}^x-S_{R}^x)+\sum\limits_{i={\rm L},{\rm R}}E_{\rm z}^iS^{z}_i.
\end{equation}
Here, ${\bf S}_{L}$ and ${\bf S}_{R}$ are spin operators belonging to the $L$ and $R$ QDs and $J_{\rm SE}$ is the magnitude of superexchange involving spin-conserving tunnel hoppings

\begin{equation}\label{eq:SupEx}
J_{\rm SE}=4 t^4 U\frac{U \left(12 \delta ^2+\epsilon ^2\right)-\delta  \left(8 \delta ^2+6 \epsilon ^2\right)}{\left(\epsilon ^2-4 \delta ^2\right)^2 \left(U-2 \delta \right) \left(U^2-\epsilon ^2\right)}.
\end{equation}
The second term in Eq. (\ref{eq:H2x2so}) is the lowest-order spin-orbit contribution to the exchange coupling, with $S_L^x$ and $S_R^x$ being the $x$-components of the spin operator corresponding to the $L$ and $R$ QD respectively.
The magnitude of the spin-orbit contribution $D$ is given by 
\begin{equation}\label{eq:SeSo}
D=\frac{2E_{\rm z}tt_{\rm so}\left(4E_{\rm z}^2-4\delta^2-\epsilon^2\right)}{16\left(E_{\rm z}^2-\delta^2\right)^2-8\left(E_{\rm z}^2+\delta^2\right)\epsilon^2+\epsilon^4}.
\end{equation}
The third term in Eq. (\ref{eq:H2x2so}) is the Zeeman energy with $S_i^z$ being the $z$-component of the spin operator corresponding to $i=L,\,R$ QD. In the process of deriving Eqs. (\ref{eq:H2x2so}-\ref{eq:SeSo}) we have neglected all terms with a power higher than $t^4$, and only kept the lowest order contribution involving spin-orbit interaction $\sim tt_{\rm so}$.

\begin{figure}[b!]
\begin{overpic}[width=0.47\textwidth]{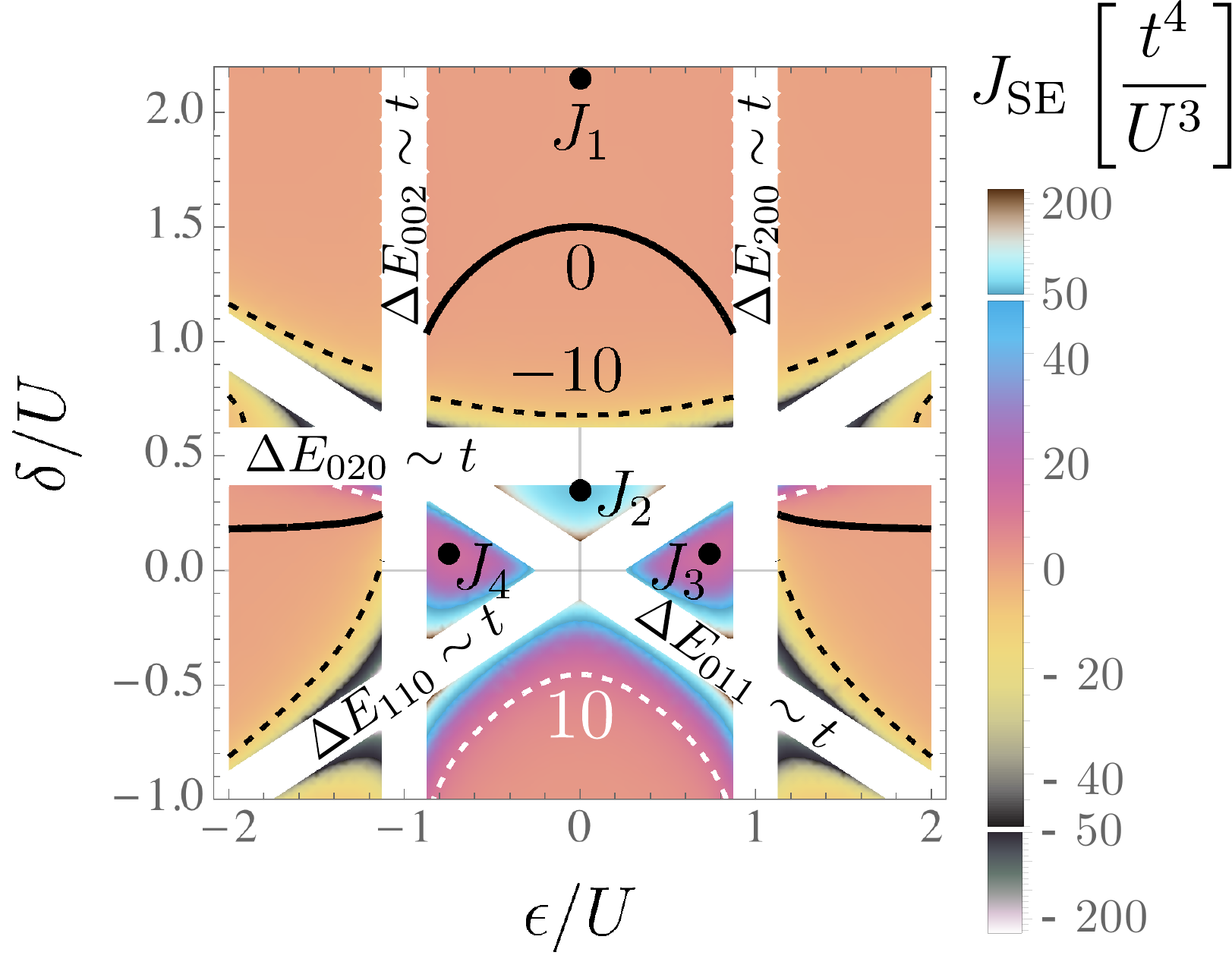}
\end{overpic}
\centering
\caption{(Color online) The superexchange in the absence of spin-orbit interaction $J_{\rm SE}$ as a function of the detuning parameters $\delta$ and $\epsilon$. The points represent the superexchange ``sweet spots" $J_1(\delta_1,\epsilon_1)=0.08$, $J_2(\delta_2,\epsilon_2)=64.65$, $J_3(\delta_3,\epsilon_3)=J_4(\delta_4,\epsilon_4)=13.62$ in units of $t^4/U^3$ where $t$ is the tunneling and $U$ is Coulomb repulsion. The black line marks $J_{\rm SE}=0$, black dashed line $J_{\rm SE}=-10$ and white dashed line $J_{\rm SE}=10$. The white regions represent areas in which the energy difference $\Delta E$ between the $(2,0,0)$ $(0,2,0)$ $(0,0,2)$ $(1,1,0)$ $(0,1,1)$ charge states and superexchange states $(101)$ becomes comparable to $t$, and therefore no superexchange takes place. Here, we chose $t=17.8\text{ $\mu$eV}$.}\label{SS}
\end{figure}

\begin{figure*}[tb!]
  \includegraphics[width=0.99\textwidth]{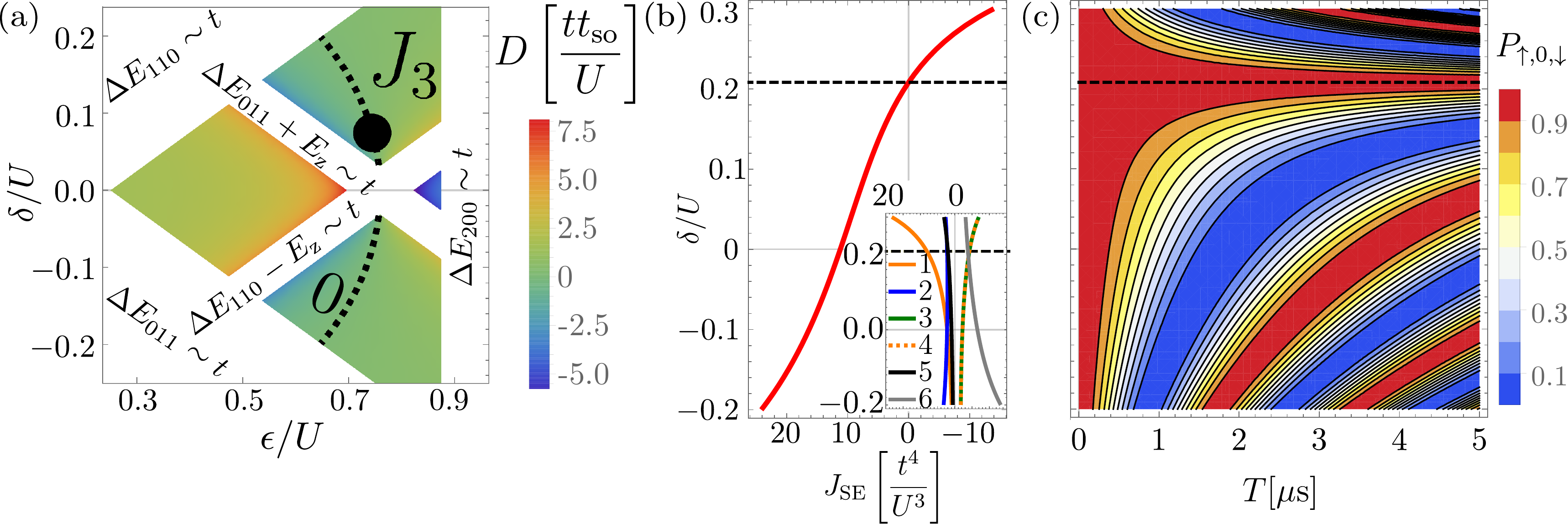}
	\caption{(Color online) (a) The strength of the spin-orbit contribution $D$ around the ``sweet spot" $J_3$ as a function of the detuning parameter $\epsilon$ and $\delta$ for $E_{\rm z}=0.38 U$. The dashed black line marks a path along which $D=0$ (b) Superexchange as a function of $\delta$ for $\epsilon=-1.34\text{ U}$ in the case of vanishing spin-orbit. Inset: magnitude of different exchange paths in the context of Tab. \ref{tab1} in the case of vanishing spin-orbit. The horizontal black dashed line represents the point $\delta_0$ in which $J_{\rm SE}=0$. (d) Coherent superexchange oscillations as a function of the detuning $\delta$ and time $T$ in the case of vanishing spin-orbit interaction. The probability to occupy the $(\downarrow,0,\uparrow)$ state is not displayed because $P_{\uparrow 0 \downarrow}=1-P_{\downarrow 0\uparrow}$. Parameters of the plots are tunneling $t=17.8\text{ $ \mu $eV}$, detuning $\epsilon=-1.34 \text{ U}$, the Coulomb repulsion $U=1\text{ meV}$.}\label{SeDyn}
\end{figure*}

A non-trivial superexchange ``sweet spot" is a point in which the superexchange is in first order insensitive to fluctuations of the detuning parameters $\delta$ and $\epsilon$, and furthermore the superexchange is not zero. Solving the coupled systems of equations $\partial J_{\rm SE}/\partial\epsilon=0$, $\partial J_{\rm SE}/\partial\delta=0$ and $J_{\rm SE}\neq 0$ for $\epsilon$ and $\delta,$ in the case of vanishing spin-orbit interaction we obtain four solutions for $\epsilon$ and $\delta$ ${\epsilon_{1,2}=0,}\,{\delta_{1,2}=(5\pm\sqrt{13})/4,}\,{\epsilon_{3,4}= \pm 0.745,}$ ${\delta_{3,4}= 0.074}$ in units of $U$ and ``sweet spots" $J_1(\delta_1,\epsilon_1)=0.08$, $J_2(\delta_2,\epsilon_2)=64.65$, $J_3(\delta_3,\epsilon_3)=J_4(\delta_4,\epsilon_4)=13.8$ in units of $t^4/U^3$ where $t$ is the tunneling and $U$ is Coulomb repulsion Fig. \ref{SS}. 

In contrast to a double QD loaded with two electrons, a linear triple QD loaded with two electrons has four points in the parameter space of $\epsilon$ and $\delta$ in which the exchange interaction is simultaneously first order insensitive in fluctuation of this two parameters. It should be noted that ``sweet spots" $J_2$, $J_3$ and $J_4$ lie close to the areas in which no superexchange takes place due to leakage outside the superexchange subspace (white regions in Fig. \ref{SS}). The width of the white areas in Fig. \ref{SS} is proportional to tunneling $t$, and this imposes a limit beyond which the magnitude of superexchange cannot by increased by increasing the tunnel coupling, while simultaneously performing superexchange at the double ``sweet spot".

We want to find values of the Zeeman energy $E_{\rm z}$ for which $D=0$ around the ``sweet spots". This would give rise to superexchange simultaneously insensitive to charge noise and spin-orbit effects in first order. By inserting $\delta_i$ and $\epsilon_i$ ($i=1,\,4$) into Eq. (\ref{eq:SeSo}) we found that such non-zero values exist corresponding to $\delta_{3,4}$ and $\epsilon_{3,4}$ and therefore to ``sweet spots" $J_{3,4}$, while no non-zero $E_{\rm z}$ for $\delta_{1,2}$ and $\epsilon_{1,2}$ exists. Two such values of the Zeeman energy exist $E_{\rm z}^{3,4}/U=\pm 0.38$ for each of the ``sweet spots" $J_{3}$ and $J_{4}$. The Coulomb repulsion $U\sim 1\text{ eV}$ in InGaAs quantum dots. The Zeeman energy of $\pm 0.38 U$ corresponds to an external magnetic field of $B_{\text{GaAs}}=\pm U/0.44\mu_B=\pm 14.9\text { T}$. However, due to a much higher $g$-factor this field is $B_{\text{InAs}}=\pm U/14.7\mu_B=\pm 0.45\text { T}$ for InAs, and thus easier to achieve. As shown in Fig. \ref{SeDyn} (a) the point $J_3$ at $E_{\rm z}=\pm 0.38 U$ is a ``super sweet spot" in which the superexchange is simultaneously insensitive to charge noise and spin-orbit effects are vanishing.
It should be noted that spin-orbit interaction is much stronger in InAs compared to GaAs. 

Solving $J_{SE}=0$ (Eq. (\ref{eq:SupEx})) we calculate $\delta_0$ for which the spin-conserving superexchange is zero for any value of $\epsilon$ and $\epsilon_0$ for which the superexchange is zero for any value of  $\delta$ see Fig. \ref{SS}.

\begin{equation}\label{eq:delta0}
\delta_0=\frac{1}{2}\left(1+\frac{1-\epsilon^2}{q^{1/3}}+q^{1/3}\right);\,\epsilon_0=\pm\frac{2\sqrt{(3-2\delta)\delta^2}}{\sqrt{6\delta-1}},
\end{equation} where, $q=1-\epsilon^2+\sqrt{(\epsilon^2(\epsilon^2-1)^2)}$ all given in units of Coulomb repulsion U.
It should be noted that the result is symmetric with respect to the sign of $\epsilon$.
When $\epsilon=-1.34U$, at large negative values of the bias $\delta$ the main contribution of the superexchange comes from the path 6 which gives rise to negative superexchange (see Tab. \ref{tab1}) as the bias is increased towards the positive values, the superexchange path 1 becomes more dominant yielding a positive sign of superexchange (see Fig. \ref{SeDyn} (b) and Fig. \ref{SeDyn} (b) inset).

Now we will investigate the dynamical evolution of spin states caused by superexchange interaction in the absence of spin-orbit interaction. We start by initializing a $(\uparrow,0,\downarrow)$ state.
The time evolution of the system in the superexchange subspace is modeled in the following way $\psi_{\rm SE}(T)=\mathcal{U}\psi_{\rm SE}(0)$, where $\psi_{\rm SE}(0)$ is the initial wavefunction corresponding 
to the initialization of the $(\uparrow,0,\downarrow)$ state, $\psi_{\rm SE}(T)$ the wavefunction at time $T$, and $\mathcal{U}=\exp{(-i\tilde{H}T/\hbar)}$ where $\tilde{H}$ is given by Eq. (\ref{eq:H2x2so}). In Fig. \ref{SeDyn} (d) we observe that superexchange oscillations are suppressed around the point $\delta=\delta_0$. Areas above and below the black line correspond to different signs of superexchange. 

\textit{Conclusion.}--We have investigated coherent superexchange and found points in parameter space in which the superexchange is both insensitive to charge noise and the spin-orbit contribution is zero. Furthermore, we have shown that the sign of the superexchange can be changed by varying the detuning parameters.
An experimental implementation of our findings would allow for charge noise-insensitive, error-free two-qubit operation of the spin $1/2$ qubit and charge-noise-insensitive, error-free control of the $S-T_0$ qubit around the exchange axis. The implications of our findings to the operation of the exchange only qubit in a charge-noise-insensitive manner are planned as a forthcoming investigation. 

\textit{Acknowledgments.}--We acknowledge funding from the EU within the Marie Curie ITN Spin-Nano and the German Research Foundation within the SFB 767.

\bibliographystyle{apsrev}
\bibliography{References4}
\pagebreak
\begin{widetext}
\section{Supplementary Material for: Low-Error Operation of Spin Qubits with Superexchange Coupling}

\section{The Schrieffer-Wolff Transformation}
\begin{figure}[b!]
	\centering
	\includegraphics[width=0.85\textwidth]{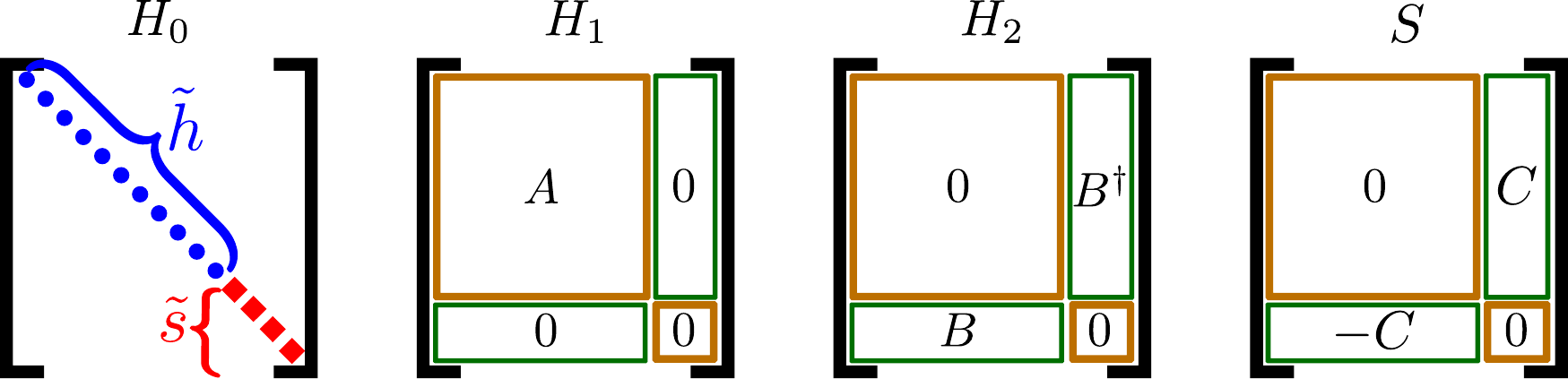}
	\caption{Schematic representation of matrices used in our Schrieffer-Wolff transformation. The full Hamiltonian is divided into the diagonal part of the Hamiltonian $H_0$ consisting of high energy $\tilde{h}$ and superexchange states $\tilde{s}$. The interacting part $H'$ is divided into $H_1$ consisting of interactions between the high energy states ($A$) and the part $H_2$ which describes the coupling between the low energy and high energy states. $S$ is an anti-Hermitian matrix which has the same block structure as $H_2$ and $C=C^{\dagger}$. It should be noted that no direct interaction between the superexchange states is present (as we assume an idealized situation with an identical $g$-factor in every dot).}
	\label{SWsup}
\end{figure}
The full Hamiltonian $H=H_0+H'$ comprises of the diagonal part $H_0$ and the off-diagonal part $H'$ 
\begin{align}
	H_0&=\sum\limits_{i\sigma}(\varepsilon_i+E_{\rm z}^i{\bf \sigma})n_{i\sigma}+U\sum\limits_i n_{i\uparrow} n_{i\downarrow}+\sum\limits_{\langle ij \rangle}V n_in_j\\
	H'&=\sum\limits_{\langle ij \rangle}\Big[\sum \limits_{\sigma} t_{ij} c_{i\sigma}^{\dagger}c_{j\sigma}+ \sum \limits_{\sigma\neq\bar{\sigma}} t_{ij}^{\rm so} c_{i\sigma}^{\dagger}c_{j\bar{\sigma}}\Big]. \label{eq:HamSup}
\end{align} Here, $E_{\rm z}$ is the Zeeman energy due to an external magnetic field, $t_{ij}^{\rm so}$ the magnitude of 
spin-non-conserving tunnel hopping caused by spin-orbit interaction, $t_{ij}$ is the magnitude of spin-conserving tunnel hopping between dots $i$ and $j$. Furthermore, $\varepsilon_i$ the energy bias of the $i$-th dot, $U$ is the Coulomb penalization of the doubly occupied quantum dot, $V$ is the Coulomb energy of two neighboring dots occupied with single electron, and $n_i=n_{i\uparrow}+n_{i\downarrow}=c_{i\uparrow}^{\dagger}c_{i\uparrow}+c_{i\downarrow}^{\dagger}c_{i\downarrow}$ the number operator, with $c_{i\sigma}(c_{i\sigma}^{\dagger})$ being the spin creation (annihilation) operator of the $i$ charge state with spin $\sigma=\{\downarrow$, $\uparrow$\}.
The $\langle ij\rangle$ in the index of the sum denotes that the runs over nearest neighbor QDs $i$ and $j$, 
and the index $\sigma\neq\bar{\sigma}$ denotes a double sum which runs over all possible possible configurations with opposite spin.
We assume a linear arrangement, neglecting all direct couplings between the $L$ and $R$ QDs. 

The Hamiltonian $H$ is $15-$dimensional and it comprises of the $11-$dimensional high energy subspace 
$\tilde{h}=
\{\textcolor{blue}{{S(2,0,0),}\,{S(0,2,0),}}\, \textcolor{blue}{{S(0,0,2),}\,{S(1,1,0),\,}{T_0(1,1,0),\,}{S(0,1,1),\,}{T_0(0,1,1),\,} {T_+(1,1,0),\,}{T_-(1,1,0),\,}{T_+(0,1,1),}\,{T_-(0,1,1)}}\}$ and the $4-$dimensional (low energy) superexchange subspace $\tilde{s}={\{\textcolor{red}{S(1,0,1),\,T_0(1,0,1),\,T_+(1,0,1),\,T_-(1,0,1)}\}}$.
Here, $S$ stands for the $m_s=0$ singlet and $T_0$, $T_+$, $T_-$ are the $m_s=0,1,-1$ triplets, respectively. Numbers in the parentheses denote charge states.
The diagonal part of the Hamiltonian $H_0$ comprises of superexchange states $\tilde{s}$ and high energy states $\tilde{h}$ (see Fig. \ref{SWsup})
\small
\begin{equation}
H_0={\rm diag}\big(\textcolor{blue}{U+\epsilon},\textcolor{blue}{U-2 \delta},\textcolor{blue}{U-\epsilon},\textcolor{blue}{\frac{\epsilon}{2}-\delta},\textcolor{blue}{\frac{\epsilon}{2}-\delta},\textcolor{blue}{-\frac{\epsilon}{2}-\delta },\textcolor{blue}{-\frac{\epsilon}{2}-\delta },\textcolor{blue}{E_{\rm z} +\frac{\epsilon}{2}-\delta},\textcolor{blue}{\frac{\epsilon}{2}-E_{\rm z}-\delta },\textcolor{blue}{E_{\rm z} -\frac{\epsilon}{2}-\delta}, \textcolor{blue}{-E_{\rm z} -\frac{\epsilon}{2}-\delta},\textcolor{red}{0},\textcolor{red}{0},\textcolor{red}{E_{\rm z}},\textcolor{red}{-E_{\rm z}}\big),
\end{equation}	
\normalsize
where the detunings $\varepsilon_i$ from Eq. (\ref{eq:HamSup}) were rewritten as $\epsilon$, the detuning between the outer dots and $\delta$ the detuning between the average of the outer dots and the middle dot.

The interaction part of the Hamiltonian $H'=H_1+H_2$ can be divided into terms containing interaction between different $\tilde{h}$ states $H_1$ and terms containing interactions between the $\tilde{s}$ and $\tilde{h}$ states $H_2$ (see Fig. \ref{SWsup})
\begin{equation}\label{eq:SupHam1}
H_1=\begin{pmatrix}
A & 0\\
0 & 0
\end{pmatrix},
\end{equation}
where $A$ is given by
\settowidth\mylen{$-\sqrt{2} t_{LC}$}
\begin{equation}
A=\left(
\begin{array}{ccccCcCCCCC}
0 & 0 & 0 & -\sqrt{2} t_{LC} & 0 & 0 & 0 &t_{LC}^{\rm so}/2 &t_{LC}^{\rm so}/2 & 0 & 0 \\
0 & 0 & 0 & -\sqrt{2} t_{LC} & 0 & -\sqrt{2} t_{CR} & 0 &t_{LC}^{\rm so}/2 &t_{LC}^{\rm so}/2 & t_{CR}^{\rm so}/2 & t_{CR}^{\rm so}/2 \\
0 & 0 & 0 & 0 & 0 & -\sqrt{2} t_{CR} & 0 & 0 & 0 & t_{CR}^{\rm so}/2 & t_{CR}^{\rm so}/2 \\
-\sqrt{2} t_{LC} & -\sqrt{2} t_{LC} & 0 & 0 & 0 & 0 & 0 & 0 & 0 & 0 & 0 \\
0 & 0 & 0 & 0 & 0 & 0 & 0 & 0 & 0 & 0 & 0 \\
0 & -\sqrt{2} t_{CR} & -\sqrt{2} t_{CR} & 0 & 0 & 0 & 0 & 0 & 0 & 0 & 0 \\
0 & 0 & 0 & 0 & 0 & 0 & 0 & 0 & 0 & 0 & 0 \\
t_{LC}^{\rm so}/2 &t_{LC}^{\rm so}/2 & 0 & 0 & 0 & 0 & 0 & 0 & 0 & 0 & 0 \\
t_{LC}^{\rm so}/2 &t_{LC}^{\rm so}/2 & 0 & 0 & 0 & 0 & 0 & 0 & 0 & 0 & 0 \\
0 & t_{CR}^{\rm so}/2 & t_{CR}^{\rm so}/2 & 0 & 0 & 0 & 0 & 0 & 0 & 0 & 0 \\
0 & t_{CR}^{\rm so}/2 & t_{CR}^{\rm so}/2 & 0 & 0 & 0 & 0 & 0 & 0 & 0 & 0 \\
\end{array}
\right),
\end{equation}
and
\begin{equation}\label{eq:SupHam2}
H_2=
\begin{pmatrix}
0 & B^{\dagger}\\
B & 0
\end{pmatrix},
\end{equation}
where $B$ is given by
\settowidth\mylen{$-t_{\rm CR}$}
\begin{equation}
B=\left(
\begin{array}{CCCcccccccc}
0 & 0 & 0 &-t_{\rm CR} & 0 & -t_{\rm LC}& 0 &-t_{CR}^{\rm so}/2 \sqrt{2} & -t_{CR}^{\rm so}/2 \sqrt{2} & -t_{LC}^{\rm so}/2 \sqrt{2} & -t_{LC}^{\rm so}/2 \sqrt{2}\\
0 & 0 & 0 & 0 & -t_{\rm CR} & 0 & -t_{\rm LC} & 0 & 0 & 0 & 0\\
0 & 0 & 0 & t_{\rm CR}^{\rm so}/2\sqrt{2} & 0 & t_{\rm LC}^{\rm so}/2\sqrt{2} & 0 & -t_{\rm CR} & 0 & -t_{\rm LC} & 0\\
0 & 0 & 0 &t_{\rm CR}^{\rm so}/2\sqrt{2} & 0 & t_{\rm LC}^{\rm so}/2\sqrt{2} & 0 & 0 & -t_{\rm CR} & 0 & -t_{\rm LC}
\end{array}
\right).
\end{equation}
Here, $t_{ij}^{\rm so}$ the magnitude of 
spin-non-conserving tunnel hopping caused by spin-orbit interaction, $t_{ij}$ is the magnitude of spin-conserving tunnel hopping between dots $i$ and $j$ (left $L$, center $C$ and right $R$).

We apply the following unitary transformation to the Hamiltonian $\tilde{H}=e^{-S}He^{S}$, where $S$ must be anti-Hermitian to ensure the unitarity of the transformation. Expanding $e^{\pm S}$ in a Taylor series the Hamiltonian equals
\begin{equation}
\tilde{H}=\sum\limits_{k=0}^{\infty}\frac{1}{k!}[H,S]^{(k)},
\end{equation}
where $[H,S]^{(k+1)}=[[H,S]^{(k)},S]$ and $[H,S]^{(0)}=H$. 
Assuming that $S$ has the same block structure as $H_2$, the transformed Hamiltonian can be separated into a block-diagonal (BD) part (having the block structure of $H_0+H_1$) and off-diagonal (OD) part (having the same structure as $H_2$)
\begin{gather}
	\tilde{H}_{OD}=\sum\limits_{k=0}^{\infty}\frac{1}{(2k+1)!}[H_0+H_1,S]^{(2k+1)}+\sum\limits_{k=0}^{\infty}\frac{1}{(2k)!}[H_2,S]^{(2k)},\nonumber\\
	\tilde{H}_{BD}=\sum\limits_{k=0}^{\infty}\frac{1}{(2k)!}[H_0+H_1,S]^{(2k)}+\sum\limits_{k=0}^{\infty}\frac{1}{(2k+1)!}[H_2,S]^{(2k+1)}.\label{eq:HDiagOffDiag}
\end{gather}
The goal of the Schrieffer-Wolff transformation is to derive the effective Hamiltonian in the (BD) form. Assuming the separation between the superexchange states $\tilde{s}$ and the high energy states $\tilde{h}$ is large compared to the tunnel couplings $|E_{\tilde{s}}-E_{\tilde{h}}|\gg t_{ij},\,t_{ij}^{\rm so}$ we can write $S=S_1+S_2+S_3...$, where each $S_k\propto t^k$ and $t\sim t_{ij},\,t_{ij}^{\rm so}$.

Every order of $S$ is determined by requiring that the OD part of the effective Hamiltonian vanishes. This gives rise to a set of coupled equations which can be iteratively solved for $S_k$
\begin{align}
	[H_0,S_1]&=-H_2,\nonumber \\
	[H_0,S_2]&=-[H_1,S_1],\nonumber \\
	[H_0,S_3]&=-[H_1,S_2]-\frac{1}{3}[[H_2,S_1],S_1].\label{eq:SSup}
\end{align}
By inserting Eq. (\ref{eq:SSup}) into Eqs. (\ref{eq:HDiagOffDiag}) we obtain the following expressions for the effective Hamiltonian in $k$th order of perturbation $\tilde{H}^{(k)}$

\begin{align}
	\tilde{H}^{(0)}&=H_0,\nonumber\\
	\tilde{H}^{(1)}&=H_1,\nonumber\\
	\tilde{H}^{(2)}&=\frac{1}{2!}[H_2,S_1],\nonumber\\
	\tilde{H}^{(3)}&=\frac{1}{2!}[H_2,S_2],\nonumber\\
	\tilde{H}^{(4)}&=\frac{1}{2!}[H_2,S_3]-\frac{1}{4}[H_2,S_1]^{(3)}.
\end{align}
This yields
\begin{equation}
\tilde{H}=\tilde{H}^{(0)}+\tilde{H}^{(1)}+\tilde{H}^{(2)}+\tilde{H}^{(3)}+\tilde{H}^{(4)}=H_0+H_1+\frac{1}{2!}[H_2,S_1+S_2+S_3] -\frac{1}{4!}[[[H_2,S_1]S_1],S_1].
\end{equation}

\section{Comparison between the evolution involving the superexchange subspace and full Hilbert space} \label{App:AppendixB}
Here, a comparison is presented between the time evolution governed by an effective $4\times4$ Hamiltonian in the $s={\{(\uparrow,0,\downarrow),\,(\downarrow,0,\uparrow),\,(\uparrow,0,\uparrow),\,(\downarrow,0,\downarrow)\}}$ subspace, obtained by eliminating $11$ states with a Schrieffer-Wolff transformation and the time evolution governed by a $15 \times 15$ Hamiltonian involving all states (Fig. \ref{vs}). It should be noted that the results presented here and in the main part of the paper are in the $s={\{(\uparrow,0,\downarrow),\,(\downarrow,0,\uparrow),\,(\uparrow,0,\uparrow),\,(\downarrow,0,\downarrow)\}}$ basis. The $\tilde{s}$ and $\tilde{h}$ bases are connected with $s$ and $h$ bases by a unitary Hadamard basis transformation.

The two ways of modeling time evolution produce results which do not differ by more then $5\%$, and therefore the result obtained by the Schrieffer-Wolff transformation are valid in the domain of applicability of the Schrieffer-Wolff transformation.

\begin{figure}[h!]
	\vspace{0.5cm}
	\begin{minipage}{0.3\textwidth}
		\includegraphics[height=4.2cm]{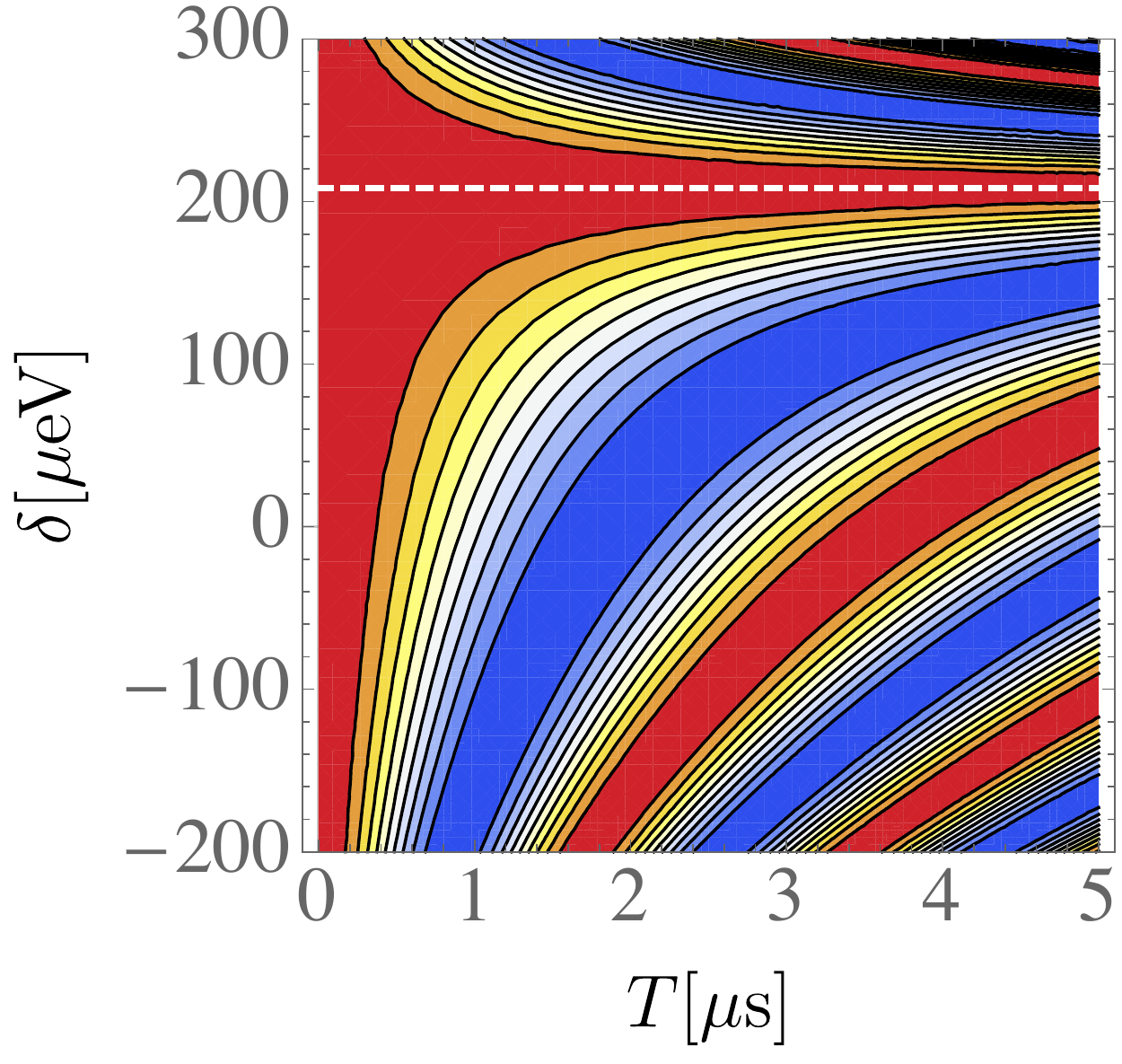}
	\end{minipage}\hspace{0.2cm}\begin{minipage}{0.3\textwidth}
		\includegraphics[height=4.2cm]{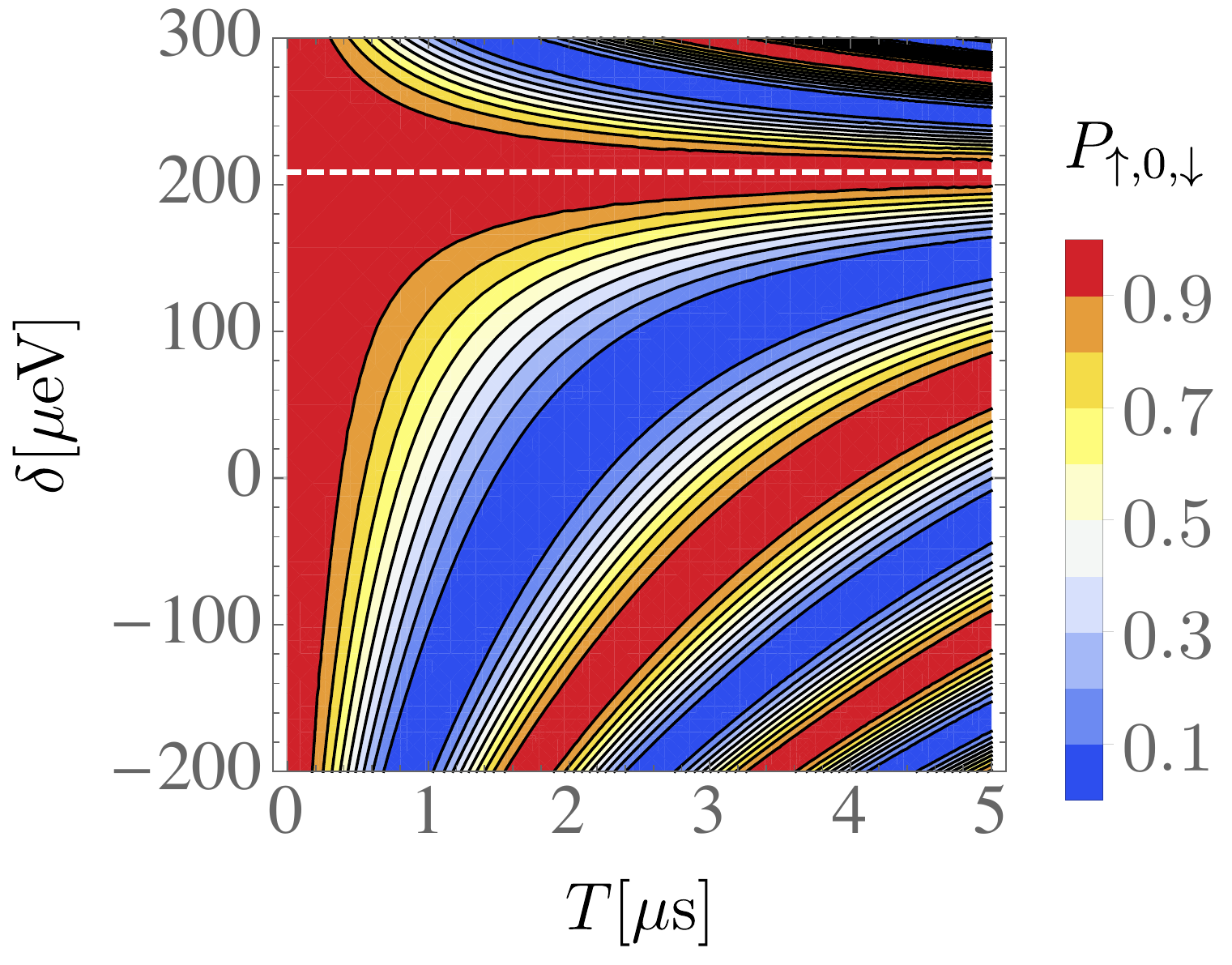}
	\end{minipage}
	\begin{minipage}{0.3\textwidth}
		\includegraphics[height=4.2cm]{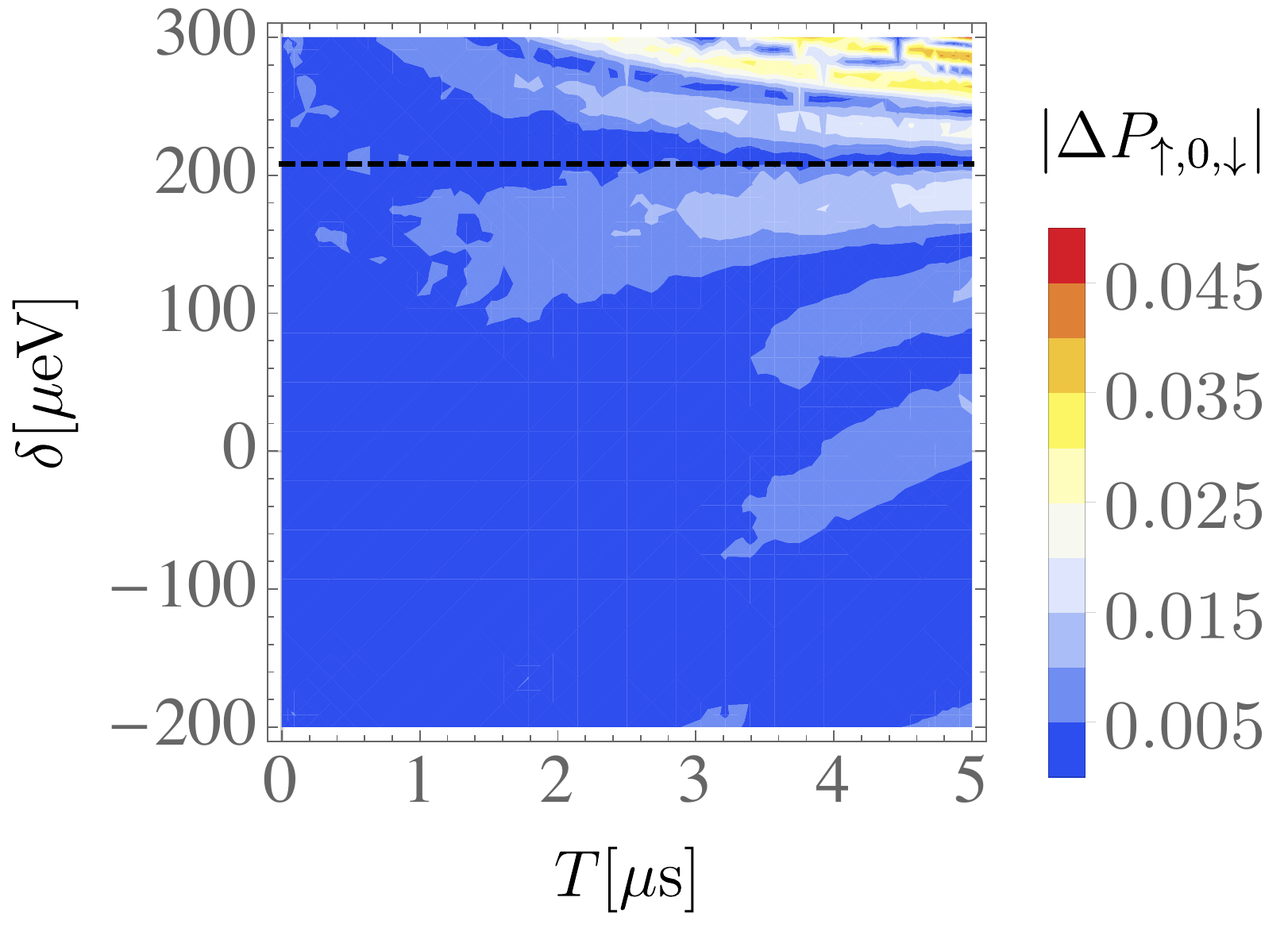}
	\end{minipage}
	\caption{Coherent superexchange oscillations of the $(\uparrow,0,\downarrow)$ state occupation probability $P_{\uparrow,0,\downarrow}$ as a function of the detuning between the middle dot and the average of outer dots $\delta$ and time $T$. The dashed line represents the point $\delta_0$ where the superexchange is zero. Parameters of the plot are the tunneling $t=17.8$  $\mu$eV, the Coulomb repulsion $U=1\text{ meV}$, and detuning between the outer dots $\epsilon=-1.34 U$. (a) The probability to occupy the $(\uparrow,0,\downarrow)$ state when all $15$ states included in the modeling of the dynamics. (b) The probability to occupy the $(\uparrow,0,\downarrow)$ state when only superexchange states are included in the modeling of the dynamics and the remaining $11$ states are eliminated with a Schrieffer-Wolff transformation. (c) The absolute difference of probabilities to occupy the $(\uparrow,0,\downarrow)$ state between the evolution when all $15$ states are included and when $11$ states are eliminated with a Schrieffer-Wolff transformation. The probability to occupy the $(\downarrow,0,\uparrow)$ state is not displayed as $P_{\uparrow 0 \downarrow}\approx 1-P_{\downarrow 0\uparrow}$.}\label{vs}  
\end{figure}

\begin{textblock}{20}(23,140)
	\textcolor{black}{(a)}
\end{textblock}
\begin{textblock}{20}(75,140)
	\textcolor{black}{(b)}
\end{textblock}
\begin{textblock}{20}(132,140)
	\textcolor{black}{(c)}
\end{textblock}
\end{widetext}
\end{document}